\documentclass[floatfix]{emulateapj}

\newcommand{\msun}{\,M_\odot}
\newcommand{\Iso}[2]{^{#1}{\rm #2}}

\shorttitle{M67 solar abundance test}
\shortauthors{Magic et al.}

\begin{document}


\title{On using the CMD morphology of M67 to test solar abundances}


\author{Z.  Magic and A.  Serenelli\altaffilmark{1} and A. Weiss}
\affil{Max-Planck-Institut f\"ur Astrophysik,
  Karl-Schwarzschild-Str.~1, 85748~Garching, Germany}
\email{magic@mpa-garching.mpg.de} 
\and
\author{B. Chaboyer}
\affil{Dartmouth College, Hanover, NH, USA}
\altaffiltext{1}{Instituto de 
    Ciencias del  Espacio (CSIC), Facultad  de Ci\`encies, Campus  UAB, 08193,
    Bellaterra, Spain}

\begin{abstract}
The open cluster M67 has solar metallicity and an age of about 4~Gyr. The
turn-off
mass is close to the minimum mass for which solar metallicity stars develop a
convective core during main sequence evolution as a result of the development
of hydrogen burning through the CNO-cycle. The morphology
of the color-magnitude-diagram (CMD) of M67 around the turn-off shows a clear
hook-like  feature,  direct  sign  that  stars  close  to  the  turn-off  have
convective cores.  
VandenBerg et al. investigated the  possibility of using the morphology of the
M67 turn-off to put constraints on the solar metallicity, particularly CNO
elements, for which solar abundances have been revised downwards by more
than 30\% over the  last few years. Here, we extend their  work filling in the
gaps in their analysis. To this aim, we compute isochrones
appropriate for M67 using new (low metallicity) and old (high metallicity) solar
abundances and study whether the characteristic turn-off in the CMD of M67 can
be reproduced or not. We also study the importance of other constitutive physics
on determining the presence of such a hook, particularly element diffusion,
overshooting  and nuclear reaction  rates. We  find that  using the  new solar
abundance 
determinations, with low CNO abundances, makes it more difficult to reproduce
the characteristic  CMD of M67.  This result is  in agreement with  results by
VandenBerg et al. However, changes in the constitutive physics
of the models, particularly overshooting, can influence and alter this result
to the extent that isochrones constructed with models using low CNO
solar abundances can also reproduce the turn-off morphology in M67.
We conclude that only if all factors affecting the
turn-off morphology are completely under control (and this is not the case), M67
could be used to put constraints on solar abundances.
\end{abstract}


\keywords{Sun: abundances --- open clusters and associations:
  individual (M67)} 



\section{Introduction}
\label{s:intro}

The latest revision of the solar abundances by \citet[AGS05]{ags:2005}
resulted in a drastic reduction of the abundances of carbon, nitrogen,
and oxygen, and thus in the total metallicity of the Sun. While $Z/X$
in   the    previous   standard   references    by   \citet[GN93]{gn:93}   and
\citet[GS98]{gs:98} 
was 0.0245 respectively 0.0230, \citet{ags:2005} determined a value of
0.0165. The recently published complete re-analysis of all elements by
\citet[AGS09]{ags:2009} corrected this number slightly upwards to 0.0183.
It is interesting to realize that $Z/X$ has constantly gone
down with time since \citet{ag:89}, who gave a value of 0.0275.

While the analysis by  \cite{ags:2005} and in particular by
\cite{ags:2009} is undoubtedly the most complete, coherent and
technically advanced one, the results are far from being accepted. The
reason is that standard solar models (SSM), calculated for the new metal
abundances, agree much less with well established and accurately
determined results from helioseismology: the sound
speed and density profile, the depth of the convective envelope, and
its helium content. \citet{bbps:2005} investigated the
new ``solar model problem'', which has been confirmed in many
subsequent, independent publications. \citet{ab:2006} further
demonstrated that helioseismic data require  the solar envelope
to have a composition close to the old photospheric abundances,
and \citet{csbev:2007} used low-degree p-modes penetrating to the
solar core to arrive at a similar conclusion.

To cure the ``trouble in paradise'' \citep{ags:2005} attempts were
undertaken to confirm the previous, higher abundances of intermediate
elements, in particular of oxygen
\citep[e.g.\ ][]{csn:2008,clsab:2008}. On the other hand, the lower
abundances were recovered, too (\citealt{mcbb:2009} for nitrogen;
\citealt{snn:2007} for oxygen).

On the side of the solar model community, possible
or necessary changes to the constitutional physics were discussed to
restore the previous excellent agreement with helioseismology (see
\citealt{guzik:2008} for a summary). In particular, an increase of
opacities has been considered the most promising approach
\citep{ba:2004,bbs:2005,ab:2006,cddmhp:2009}. But updated Rosseland
opacities for the  solar interior are not enough to  solve the solar abundance
problem  \citep{bsb:2005}. Also, a  postulated upward  correction to  the neon
abundance has not found observational support \citep{young:2005}.

It is therefore reasonable to shift attention also to other stellar
objects, since any revision of the solar metallicity yardstick leads
to a corresponding change in the metallicity of all stars. This is in
particular true for stars with $\mathrm{[Fe/H]}\approx 0$, since their
metallicity is often determined in a strictly differential and
therefore quite accurate way. \citet{algdc:2007} found that with the
lower AGS05 solar composition they could reproduce the properties of
the pre-main sequence binary system RS~Cha, while it was impossible to
fulfill all observational constraints with the higher GN93
abundances.

Another obvious test was performed by \citet[VG07]{vgeef:2007}, using
the CMD morphology of the open cluster M67, which has a metallicity of
$\mathrm{[Fe/H]}= 0.00\pm0.03$ \citep{rsppp:2006} and a turn-off (TO) age
around  4~Gyr \citep[e.g.\ ][]{pietr:04}. At this age the turn-off mass
is slightly higher than $1\msun$ and close to the critical mass at
which a convective core on the main-sequence appears due to the
dominance of the CNO-cycle over the pp-chain \citep{kw:90}. The
CMD of M67 \citep{mmm67:93,esm67:2004} shows a clear hook-like
structure at the turn-off, indicative of a convective core for the
turn-off stars. Since the efficiency of the CNO-cycle is directly
proportional to the abundance of these elements, the reduced solar
abundances of AGS05 could possibly result in a radiative core for
stars at the TO, although in the fitting process the change in
luminosity  (at given mass,  luminosity is  lower for  the AGS05  mixture) and
effective 
temperature complicates simple predictions about the TO-mass and core
structure. 
Indeed, VG07 found that the TO-hook disappeared for
the new abundances, thereby strengthening the case for the GS98
metallicity scale. 

However, VG07 did not fail to mention some caveats: Their models were
calculated without atomic diffusion taken into account, a physical
process that at least for solar models -- irrespective of the
metallicity scale used -- is essential for the best possible agreement with
seismic inferences. They correctly pointed out that diffusion helps to
support a convective core at a stellar mass lower than in models
ignoring this effect \citep{mrrv:2004}, mainly due to the increase of
CNO abundances in the stellar core by gravitational settling. 
The different  ages of M67 and  the Sun, but  identical present-day abundances
could also 
imply different initial abundances, an effect that should be taken into
account in careful and precise studies. Apart from these points, one
should keep in mind that with the occurrence of a convective core the
question about the amount of overshooting arises. Its treatment, and 
in particular the value of any free parameter in its practical
implementation, cannot simply be taken 
for granted, as usually such ``calibration'' was based on stellar
models resting on the higher solar metallicity scale. 
Finally, the occurrence of a convective core may also depend on other
aspects and therefore any conclusion concerning the metallicity is
valid only under the assumption that all other influences are under
control. 

Given the result by VG07 and the open questions raised above, we revisited
M67, trying to complete and extend the pioneering study of VG07. In
\S~\ref{s:moddat} we will introduce briefly the stellar evolution
code we have 
used for  most of the calculations in  the present work, and  those aspects of
the input physics, besides the solar metallicity, we will also investigate. In
\S~\ref{s:fits} we demonstrate that we can recover the results of 
VG07, a fact that is not irrelevant in the light of the following
section. We then present our preferred model, showing that in this case
the CMD morphology of M67 can be reproduced for both solar abundance
scales. In \S~\ref{s:fitn} we investigate, how other
approaches and variations of the input physics may also influence the
quality of  isochrone fits, and  also possible systematic  differences arising
from the use of different stellar evolution codes. Our conclusions will
follow in \S~\ref{s:conc}.

\section{Stellar models and crucial data}
\label{s:moddat}

\subsection{GARSTEC}
\label{s:garstec}

Most of the stellar model calculations  presented in this paper were done with
the GARSTEC \citep{wsch:2008} program. 
Here, we briefly present aspects or modifications of the code, which
are relevant for this work. The standard assumptions and settings can
be found in the quoted reference. 

Convective overshooting is treated as a diffusive process according to the 
approach described by \citet{freyt:96}. The exact implementation can be found 
in \citet{bloe:97}. Based on hydrodynamical simulations, the diffusive constant 
therein ($D_\mathrm{os}$) is assumed to decay exponentially beyond the formal
Schwarzschild-border as 
\begin{equation}\label{eqn:ove}
 D_\mathrm{os}\left(z\right)  =  D_0  \  \exp  \  \frac{-2z}{f  H_P},\  D_0  =
 \frac{1}{3} 
 v_0 \cdot H_P . 
\end{equation}
This is equivalent to exponentially declining velocities of 
convective elements.
In (\ref{eqn:ove}) $z$ is the radial distance from the Schwarzschild border
and $H_P$ the pressure scale height taken there.
The constant $D_0$ sets the scale of diffusive speed, and depends on
the convective velocity $v_0$ inside of the convective border.
$f$ is a free parameter defining the scale of the overshooting. It is known
that fitting the CMD of young open clusters usually leads to an overshooting
region extending for about 0.2~$H_P$ in the classical local prescription. We
have computed some test models for stars between 2 and 6~$\msun$ and found
that this overshooting value is well reproduced by the diffusive approach,
during the hydrogen core burning phase, with a value of about $f=0.018$, a
confirmation of previous results by \citet{bloe:97}.

For small convective cores, the amount of overshooting has to be
limited. In the local description this is done, for example, by a 
multiplicative factor, which increases linearly with mass in the range
of 1 to $2\, M_\odot$ \citep[see][for an example]{pietr:04}. 
In our code this is achieved by a geometrical cutoff, where $H_P$ in
Equation~\ref{eqn:ove} is replaced by 
\begin{equation}\label{eqn:ove2}
\widetilde{H_P} := H_P \cdot  \min \left[1,\left( \frac{\Delta
    R_\mathrm{cz}}{H_P}\right)^2 \right],  
\end{equation}
where $\Delta R_\mathrm{cz}$ is the thickness of the convective
zone.  In  this  way,  the   geometric  cutoff  has  the  advantage  that  the
overshooting region  is always limited to  a fraction of the  extension of the
convective region. It is our standard limiting procedure for overshooting.
A similar cut is used in \citet{vcs:2005,vzmd:98}. 

An  alternative approach,  namely  the use  of  a ramp  function  for the  $f$
parameter, is also applied for some  of our calculations.  For masses lower or
equal than 
$1.1\, M_\odot$ we use $f=0$  (i.e. no overshooting), whereas for masses equal
to or higher  than $2.0\, M_\odot$ we set $f=  0.018$. In the intermediate mass
range the efficiency of overshooting varies linearly with mass 
\begin{equation}\label{eqn:ove3}
f = (0.13M/M_{\odot} - 0.098)/9.
\end{equation}
The small  discontinuity in the  overshooting efficiency is of  no consequence
for  our study. $1.1\,  M_\odot$  models with
$f=0.005$,  as would result  from application  of the  above equation  to this
mass, lead,  given the low  overshooting efficiency, to the  same evolutionary
tracks as models without overshooting. 
This prescription for the overshooting efficiency leads to evolutionary tracks 
that reproduce well those by \citet{pietr:04}. 

Atomic diffusion  is treated  within the same  diffusive numerical  scheme. In
this paper, 
hydrogen, helium and heavier elements (including iron) are diffusing. Radiative
levitation is not taken into account.

To agree with the physical assumptions in VG07, the default set of
nuclear reaction rates are those of the NACRE collaboration
\citep{ang:99}. To investigate their influence, for 
individual reactions alternative rates were used.  For the $^{14} \mathrm
N(p,\gamma)^{15} \mathrm O$ reaction, we also used the rate from 
\citet{mfn14:2008}, the  newest result from the LUNA  collaboration, which is
lower by about a factor of two  with respect to the NACRE rate at the relevant
temperatures. This has a considerable effect on the TO morphology. 
Another reaction that turned out to be of surprisingly strong influence is
the $^{17} \mathrm O(p,\alpha)^{14} \mathrm N$ reaction. 
We tested this by employing either the recent measurements by
\citet{moaz:07}, which for $T > 4\times 10^8$ K is very similar to the NACRE
recommendation, or the one given by \citet{cf:1988}. 
A detailed discussion on the reaction rates can be found in
\S~\ref{s:nucleo}. 

For all solar composition choices (AGS05, GS98) consistent Rosseland
mean opacity tables were prepared following the procedure described in
\citet{wsch:2008}.
 
The final step to compare isochrones with observed CMDs is the
transformation  to colors.  For  this we  used  that by  \citet[VC03]{vc:03}
which, together with a choice for the distance 
modulus and reddening of M67, results in satisfying CMD fits on the
main sequence and subgiant branch. Alternatively, we used the
transformations by \citet{cas:04} for testing purposes.

The initial stellar parameters ($Y_\mathrm{in}$, $Z_\mathrm{in}$, mixing length
parameter) are obtained from solar model calibrations. This will be
discussed in more detail in  \S~\ref{s:calib}. With these, we computed the
evolution from the zero-age main-sequence (ZAMS) to the tip of the Red
Giant Branch (RGB) for mass values from $0.6$ to $1.5\ M_\odot$ in
steps of $0.1\ M_\odot$. 
We recall here that, as in VG07, we are not interested in considering the RGB
of M67, where deficiencies in our treatment of the outer layers of the star
could have an impact on the location of the models in the HRD. 
We confirmed by tests starting from the
pre-main sequence  that for stars  with mass below  the critical mass  for the
occurrence of a convective core, $M_\mathrm{ccc}$, the
transient convective core at the end of the pre-main sequence phase also
appears here. This convective core is the result of the short phase of
CN-conversion, but may be sustained if convective overshooting is
included.  Finally,  for  constructing  isochrones,  the  tracks  are
normalized to the so-called equivalent points \citep{vb:92,pietr:04}
and the interpolation to an isochrone is done between the normalized tracks.
After an isochrone age was fixed,  we calculated an additional model with the
TO-mass and  recalculated the  isochrone to make  sure that  the TO-morphology
does not depend on the isochrone interpolation scheme.

\subsection{Data for M67}
\label{s:data}

To compare our models with M67 we used the photometric data by
\citet{esm67:2004}. These are more accurate than the older data from
\citet{mmm67:93}, which were used by VG07, and contain bona fide
single stars only. Thus, this CMD has a narrower main-sequence band.
However, we also tested
some of our isochrones with the data by \citet{mmm67:93}. 
We used both $(B-V)$ and $(V-I)$ colors. 

The dereddened distance modulus to M67 is $(m-M)_V = 9.70$
according to VG07, and $9.72\pm 0.05$ \citep{esm67:2004} based on
subdwarf fitting to the lower main-sequence following \citet{psk:03}. 
The reddening is $E(B-V)=0.038$ (VG07) in good
agreement with \citet{shkd:99}, who gave $0.04\pm0.02$.
Similar values have been recently obtained by \citet{twarog:2009}.
For the metallicity we assumed a value of $\mathrm{[Fe/H]}=0.0$, which
is well within spectroscopically determined errors, for example by
\citet{gm67:2000}, who gave $\mathrm{[Fe/H]}=0.02\pm0.06$.

\subsection{Model composition}
\label{s:compos}

To be consistent with VG07, we used for the old respectively new
solar abundances those by GS98 and AGS05, although the former are
simply an update of GN93 taking into account additional literature
values, and AGS09 would be the most recent and complete
re-analysis. Since AGS09 abundances are slightly higher than AGS05,
our choice is testing the more extreme case. The initial abundances of
the stellar models for the M67 isochrones are taken to be the same as
those resulting from solar model calibrations.

\section{Standard fitting procedures}
\label{s:fits}

\subsection{Recovering VG07}
\label{s:repro}

The idea of VG07 to use M67 for testing the effect of the new solar
abundances rests on the fact that the turn-off mass of this open
cluster is very close to the critical mass for the onset of core
convection, $M_\mathrm{ccc}$, or ''transition mass''
\citep{vgeef:2007}. This transition mass depends on the average
exponent of the energy generation rate in the core
\citep[Chap.~22]{kw:90}, that increases with the contribution from the
CNO-cycle, for which $\epsilon_\mathrm{CNO} \sim T^{15}$, in contrast to
$\epsilon_\mathrm{pp} \sim T^5$ for the pp-chains at the relevant temperatures.
The importance of the CNO-cycle
depends on the nuclear reaction rates (see \S~\ref{s:nucleo}) as well
as on the amount of ``catalysts'', i.e.\ the sum of the
CNO-abundances. These are, incidentally,
the elements with the largest reduction in their abundance according
to AGS05. 
Therefore, the morphological change in the CMD,
i.e.\ displaying the characteristic hook at the TO, or a gap in
star density, can be used to determine whether the TO-mass is below or
above $M_\mathrm{ccc}$. 

However, the sensitivity of the M67 TO morphology implies that also
other effects may influence it, apart from the metallicity. The
nuclear reaction rates were already mentioned. Other possible aspects
could be the amount of overshooting, atomic diffusion, and the pre-main sequence
history. Last, but not least, technical details of the stellar
evolution codes may play a role. It is therefore necessary to show
that our results do not depend on the particular numerical code. As a
first, and crucial step, we have successfully attempted to reproduce
the key result by VG07. This will also show where changes to their
procedure are indicated.

\begin{figure}
\centerline{\includegraphics[scale=0.65]{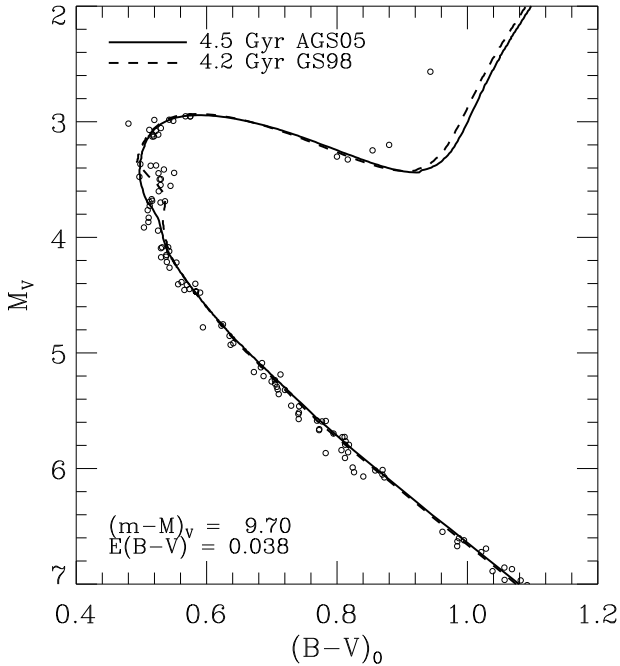}}
\caption{Isochrone fit to M67 for the two solar mixtures GS98 and
  AGS05 using a solar model calibration as in VG07 (see
  Table~\ref{t:ssm}) and NACRE nuclear reaction rates. Isochrone ages are
indicated; distance modulus and reddening agree with VG07}
\protect\label{f:vgrep}
\end{figure}

The first step concerns the solar model calibration. Here, as for the
M67 models, VG07 used NACRE nuclear reaction rates, the OPAL
\citep{ri:96} and \citet{af:05} opacities (as is done in our code),
but ignored diffusion. This point is crucial, as
we will show later on. Some amount of convective overshooting (see
\S~\ref{s:over}) was 
included for the M67 models, but is not relevant for the solar
calibration. For the atmospheres they used the MARCS model 
\citep{marcs:2003}. In this respect we differ, since we use standard
Eddington grey atmospheres. However, as shown by \citet{veeg:2008},
this has in our present case no significant influence on the tracks on
the main-sequence and subgiant branch.

\begin{table}
\caption{Results of solar calibrations for different codes and solar
  abundances (cols.~1 and 2)
  and for cases with and without diffusion (col.~3). $Y_\mathrm{in}$
  and $Z_\mathrm{in}$ are the initial helium and metal abundances. $\Delta$ 
denotes the relative difference with respect to VG07 for the
corresponding abundance.}
\label{t:ssm}
\begin{centering}
\begin{tabular}{cccccccc}
Code & Mixture & Diffusion & $\alpha_{\mathrm{MLT}}$ & $Y_ \mathrm{in}$ & 
$\Delta$ & $Z_ \mathrm{in}$ & $\Delta$  \tabularnewline
\hline
\hline 
VG07 & GS98 & no & 1.84 & 0.26760 & & 0.01650\tabularnewline
 & AGS05 & no & 1.80 & 0.25590 & & 0.01250\tabularnewline
 \hline 
GARSTEC & GS98 & no & 1.59  & 0.26109 & -2\% & 0.01661 & 1\% \tabularnewline
 & AGS05 & no & 1.62 & 0.25018 & -2\% &  0.01217 & -3\% \tabularnewline
 & GS98 & yes & 1.72  & 0.26903 & 1\% & 0.01866 & 13\%\tabularnewline
 & AGS05 & yes & 1.75 & 0.25939 & 1\% & 0.01380 & 10\%\tabularnewline
\hline 
Dartmouth & GS98 & no & 1.78 & 0.26461 & -1\% & 0.01659 & 1\% \tabularnewline
 & AGS05 & no & 1.77  & 0.24991 & -2\% & 0.01219 & -3\% \tabularnewline
 & GS98 & yes & 1.94 & 0.27419 & 2\% & 0.01889 & 14\%\tabularnewline
 & AGS05 & yes & 1.91 & 0.26075 & 2\% & 0.01405 & 12\%\tabularnewline
\hline 
LP & GS98 & no & 1.71 & 0.26466 & -1\% & 0.01648 & -0.1\% \tabularnewline
 & AGS05 & no & 1.66 & 0.25026 & -2\% & 0.01217 & -3\% \tabularnewline
 & GS98 & yes & 1.85 & 0.27053 & -1\% & 0.01806 & 10\%\tabularnewline
 & AGS05 & yes & 1.79 & 0.25728 & 1\% & 0.01340 & 7\%\tabularnewline
\end{tabular}
\end{centering}
\end{table}

Table~\ref{t:ssm} contains in the first two rows the resulting solar
model parameters by VG07 and in rows 3 and 4 the equivalent ones
obtained with our code. Additionally, we also show the results when
using the Dartmouth and LPCODE codes (\S~\ref{s:codecomp}).  Note that the 
mixing length 
parameters can never be compared in their absolute values due to the
different formulations of MLT. Overall, the agreement with VG07 for
the initial abundances is within 1--2\%, with our codes returning
systematically lower initial metallicities (by 3\%) for the AGS05
mixture. Although a small effect,  it additionally disfavors the appearance of
a convective core in the M67 TO.

Using these initial composition values we calculated stellar tracks
and isochrones for M67, which we show in Figure~\ref{f:vgrep}. The
numerical values for distance and reddening are identical to VG07. The
best-fitting isochrones have somewhat higher ages (by 0.3~Gyr). The TO
mass is $1.229$ respectively $1.200\,\msun$ for the old and new
composition. 
One recognizes the same basic 
result as in VG07: for the AGS05 mixture, the isochrone does not show
the characteristic hook. However, it displays 
a slight inclination indicating the presence of a very small
convective core at the TO mass. Consistent with this morphological
sign, the TO mass is marginally higher than $M_\mathrm{ccc}$, which is
$1.175\,\msun$, while in the case of the GS98 mixture it is clearly
above $M_\mathrm{ccc}=1.139\,\msun$. These values appear to be very
similar to those found 
by VG07 (see their Table~1). This exercise 
demonstrates that we are able to reproduce correctly VG07 and that
their result is independent of the stellar evolution code used.

\subsection{Inclusion of diffusion}
\label{s:calib}

VG07 already emphasized that atomic diffusion (effectively
sedimentation) could modify this result as it leads to an increase of
metallicity in the core over time. As explained in the previous
subsection, this will favor the occurrence of a convective core and
reduce $M_\mathrm{ccc}$. However, the solar model calibration should
also include diffusion, as it was shown that only with this physical
effect the best agreement with the seismic Sun can be achieved. Only
in this case the solar parameters, in particular the initial
composition is determined as accurately as possible. 

The corresponding results are listed in Table~\ref{t:ssm} as
well. Obviously, both helium and in particular metallicity are now much
higher as compared to the previous case ignoring diffusion in the
calibration; the latter values are increased by 10--13\% (similar
values for the comparison codes). We are using the same solar initial
abundances for the M67 isochrones. This is not completely consistent
as M67 displays {\em now} a solar metallicity, but due to its slightly
younger age diffusion should have started with a somewhat lower
metallicity than the Sun. However, this effect amounts to a few parts
in $10^{-4}$ in $Z$ only, such that we did not iterate further the
initial composition.

\begin{figure}[h]
\centerline{\includegraphics[scale=0.65]{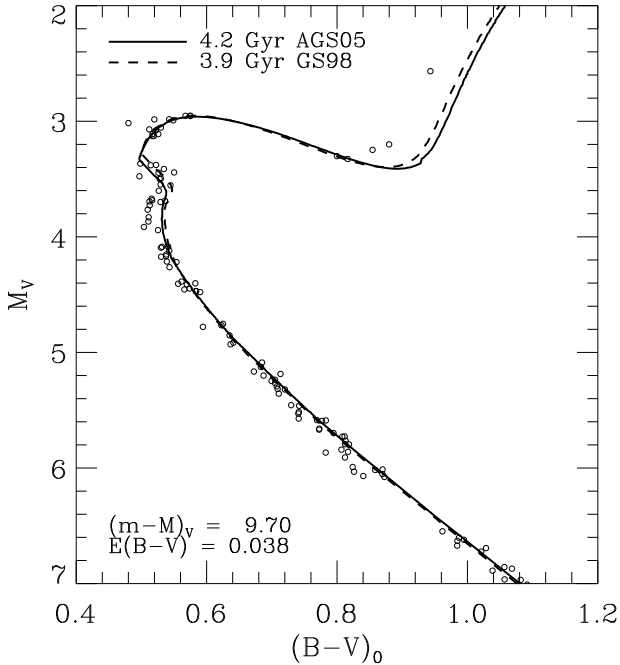}}
\caption{Isochrone fit to M67 using an initial composition obtained
  from standard solar model calibration with diffusion. Tracks for M67 include
diffusion and NACRE rates; no overshooting was considered.
\label{f:ssmdiff}}
\end{figure}

The final fits to M67 are shown in Figure~\ref{f:ssmdiff}. We still used
the same distance and reddening values, but the isochrone ages are now
lower by 0.3~Gyr compared to Figure~\ref{f:vgrep}. The agreement with
the ages by VG07 is purely incidental. Instead, the age for
the GS98 mixture of 3.9~Gyr should be compared to that by
\citet{mrrv:2004} of 3.7~Gyr. Now, for both
compositions the CMD morphology is reproduced, and AGS05 is no longer
disfavored. The fit itself is of the same quality as before.

\section{Variations}
\label{s:fitn}

In  this  section  we  investigate  the importance  of  further  physical  and
numerical aspects that influence the  appearance and size of a convective core
on the  main sequence. These are  obviously the nuclear reaction  rates of the
CNO-cycle  and the  amount  of  overshooting.  Overshooting  may  lead to  the
persistence of the otherwise transient convective core that may appear towards
the end  of the  pre-main sequence, but  may be  kept alive due  to additional
mixing of CNO-nuclei by overshooting. With this aspect we start our tests. 

\subsection{Overshooting}
\label{s:over} 

\begin{figure} 
\centerline{\includegraphics[scale=0.65]{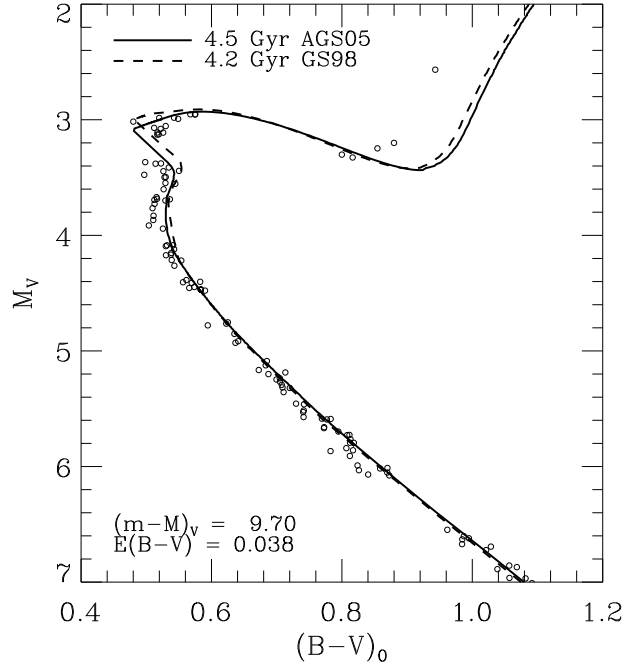}}
\caption{As Fig.~\ref{f:vgrep}, but here overshooting with a parameter 
of $f=0.018$ and geometric cutoff was used.}
\label{f:ov018}
\end{figure}

We  first  repeated  the   models  of  \S~\ref{s:repro},  i.e.  diffusion  was
completely  ignored,  but  included  overshooting according  to  our  standard
prescription             outlined            in            \S~\ref{s:garstec},
Equations~\ref{eqn:ove}~and~\ref{eqn:ove2}.  The parameter  $f$ can be varied,
but we keep it below the standard value of 0.02 \citep{bloe:97}. 
In this work we take $f=0.018$ because with this value, in our 
alternative approach (Equation~\ref{eqn:ove3}), we are able to reproduce well
the evolutionary tracks from \citet{pietr:04}, as mentioned
in \S~\ref{s:garstec}.
In Figure~\ref{f:ov018} the resulting CMD for $f=0.018$ and the standard 
cluster parameters is displayed. We recall that the amount of
overshooting  is  reduced   due  to  our   geometrical  restriction
(equation~\ref{eqn:ove2}); this  reduction can be quite  restrictive for stars
with 
lower masses,  around $1.2\,M_\odot$, and  be crucial in this  study. However,
even if the fit  itself is not as good as before,  now both mixture cases have
TO masses with a convective core being present. 

The situation seems to be typical for cases without diffusion in
both the solar calibration and the cluster models.
In \citet{vdbs:2004} the authors concluded that a small 
amount of overshooting is needed to reproduce best the CMD of
M67. Overshooting was treated following the integral criterion by
\citet{rox:89} and the adaption of a varying
amount of overshooting for very small convective cores close to
$M_\mathrm{ccc}$,   determined  by  \citet{vbd:2006}. We stress the fact that
this procedure was calibrated using open clusters with solar abundances on
the {\em old} GS98 scale. For complete consistency, this procedure should be
repeated for the new AGS05 composition, and this may explain why the
same small amount of overshooting used by VG07 did not lead to a persistent
convective core in case of the new abundances.

The case of \S~\ref{s:calib}, i.e.\ the one with full consideration of
diffusion,
was also extended by including overshooting in the tracks for M67. We
refrain from showing the result here, as for both mixtures the
convective core, as expected, persisted, but the overall fit got
worse. This coincides with the isochrone fit shown by
\citet{pietr:04}, where the CMD could be better
reproduced if overshooting was completely ignored. It agrees also
with \citet{mrrv:2004}, who neither found any necessity for
overshooting, and who, too, did include diffusion in their models.

\subsection{Nuclear reaction rates}
\label{s:nucleo}

As mentioned above, the nuclear reaction rates of the CNO-cycle are
equally important for the occurrence of a convective core as is the
abundance of CNO-nuclei. So far we have presented models employing the
NACRE-library reaction rates for these reactions. The bottleneck
reaction of the CNO-cycle, $\Iso{14}{N}(p,\gamma)\Iso{15}{O}$, that
determines the overall cycle rate, has been measured at stellar
energies in the laboratory by the LUNA collaboration
\citep{lunan14:04,mfn14:2008}, and  was found to  be lower by about  50\% with
respect to the NACRE rate. The
consequences for globular cluster age determinations and for some
sensitive phases of low- and intermediate-mass star evolution has been
investigated by \citet{icfbb:04} and \citet{wskschcd:2005}. 

\begin{figure} 
\centerline{\includegraphics[scale=0.65]{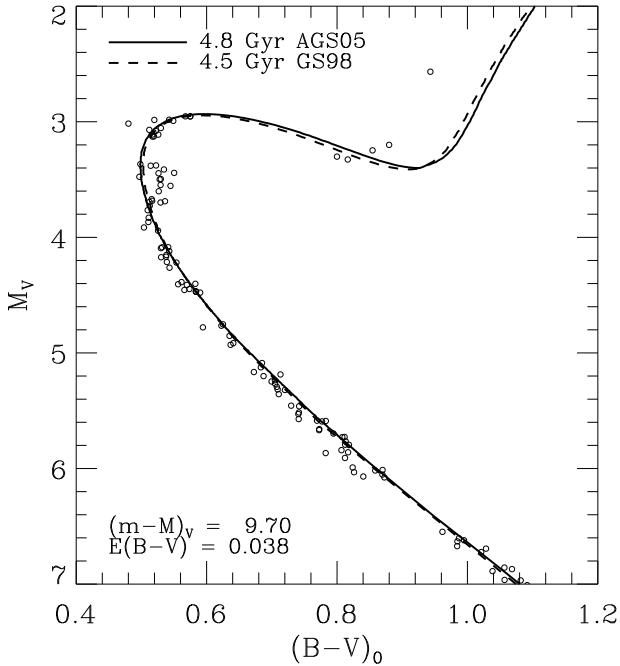}}
\caption{CMD of M67 with isochrones resulting from tracks computed with
the $\Iso{14}{N}(p,\gamma)\Iso{15}{O}$ reaction rate by
\citet{mfn14:2008} instead of that in the NACRE-library. In both the solar
calibration and these tracks diffusion and overshooting were ignored.
\label{f:vg07n14}}
\end{figure}

The rate being lower, we expect that the transition mass to harboring
a convective core increases. Therefore isochrones using the lower
CNO-abundances of the AGS05 mixtures will be even less likely to show
the TO hook. We repeated the case of \S~\ref{s:repro} (i.e.\ ignoring
diffusion completely) with the updated and most likely more accurate
rate by \citet{mfn14:2008}. The resulting CMD for M67, using again our 
standard distance 
modulus and reddening is shown in Figure~\ref{f:vg07n14}. In agreement
with  \citet{icfbb:04} the isochrone age had to be increased (by
0.3~Gyr for both mixtures) and $M_\mathrm{ccc}$ increased by $\approx
0.08\,M_\odot$ in both cases. As a consequence also the GS98 case now
lacks the characteristic hook. We add briefly that the inclusion of
overshooting  does not  alter  this  result because  the  geometric cutoff  is
restrictive  enough  that the  convective  region formed  at  the  end of  the
pre-main sequence phase  can not be maintained  during the main
sequence evolution.  The values for $M_\mathrm{ccc}$ are 1.215 (GS98) and
1.258 (AGS05), those at the TO 1.202 resp.\ 1.196~$M_\odot$. 

\begin{figure} 
\centerline{\includegraphics[scale=0.62]{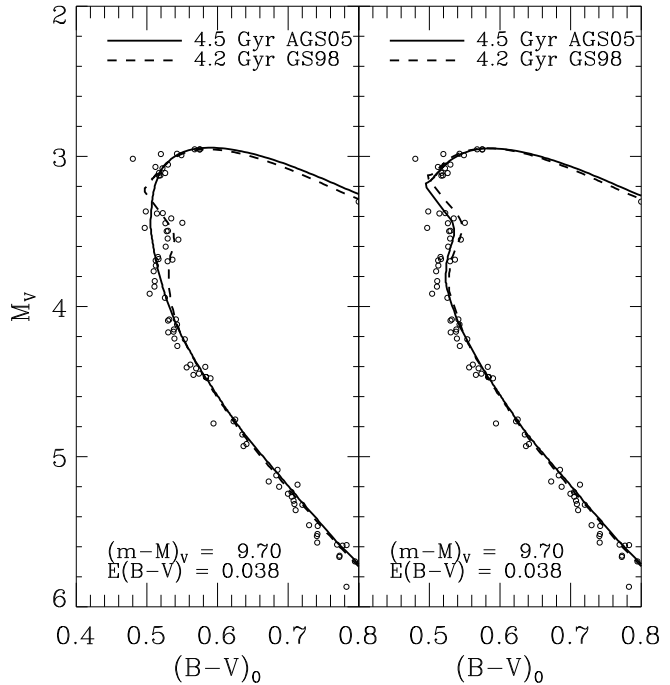}}
\caption{As Figure~\ref{f:vg07n14}, but diffusion was included in both
  the solar calibration and in the tracks for the M67 isochrones.  Left panel:
  overshooting constrained by the geometric cutoff 
(Equation~\ref{eqn:ove2}); right panel: overshooting constrained by a linear 
  dependence on stellar mass (Equation~\ref{eqn:ove3}).} 
\label{f:calibn14}
\end{figure}

This illustrates the fundamental problem with such tests: Imagine we
would assume that the GS98 solar composition is the correct one, but we
want to test which reaction rate is to be preferred. From
Figures.~\ref{f:vgrep} and \ref{f:vg07n14} we would clearly conclude that
the older one is to be preferred!

We now repeat the computation of \S~\ref{s:calib}, i.e.\ the case with
diffusion included, but with the newer and lower $\Iso{14}{N}(p,\gamma)
\Iso{15}{O}$ reaction rate. 
In this case we recognize that with the older solar composition a small 
convective core is present (TO-mass/$M_\mathrm{ccc}$ = 1.214/1.172), 
which is completely absent in the AGS05 case (1.201/1.241~$M_\odot$). 
While the CMD fit is not very good even with the GS98 isochrone, 
it  can  be  improved  by  including  overshooting.   The  results,  including
overshooting with the geometric cut, are shown on the left panel of 
Figure~\ref{f:calibn14}. As mentioned in  the previous
section, the geometric cutoff limit for 
overshooting is very restrictive particularly  for lower masses. Therefore we 
recompute  the  tracks  with  the  alternative limiting  algorithm,  the  ramp
function described by Equation~\ref{eqn:ove3}. 
Changes in TO morphology for the GS98 composition are only minor. On the other
hand, for the AGS05 composition, we now get a similarly good CMD fit as with
the old  mixture (right panel  of Figure~\ref{f:calibn14}), as  the convective
core  that appears  in  the pre-main  sequence  can be  maintained during  main
sequence evolution.  

\begin{figure} 
\centerline{\includegraphics[scale=0.65]{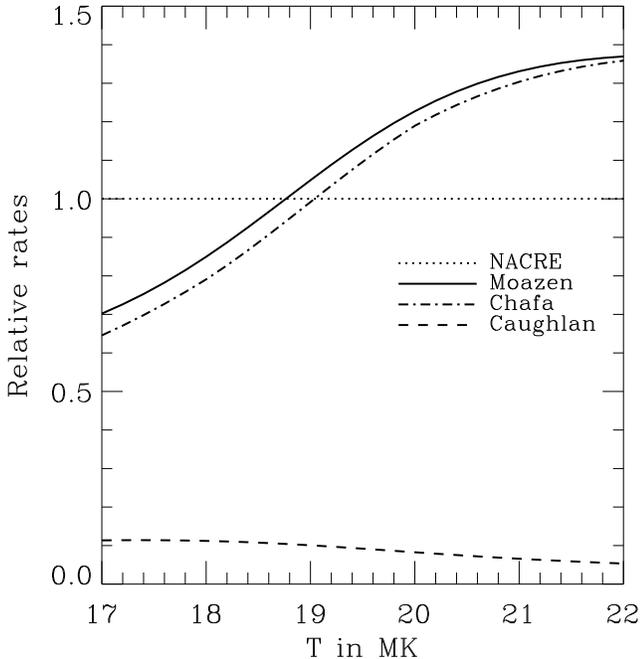}}
\caption{Rates for the nuclear reaction
  $\Iso{17}{O}(p,\alpha)\Iso{14}{N}$ relative to that of the NACRE
  library, in the interesting temperature regime for core hydrogen
  burning in typical M67 stars near the TO mass.}
\label{f:o17rate}
\end{figure}

We found a second reaction that is also influencing the efficiency of
the CNO-double cycle, though indirectly. This is the
$\Iso{17}{O}(p,\alpha)\Iso{14}{N}$ reaction that closes the CNO-II
cycle and which is, together with the competing
$\Iso{17}{O}(p,\gamma)\Iso{18}{F}$  the slowest one of that
subcycle. For this reaction new cross section measurements by
\citet{moaz:07} and \citet{chafa:07} are available. The resulting two new
reaction rates are quite similar to each other and vary in the interesting
temperature range of 15 -- 22~MK between  50\%  and 120\% of the NACRE
rate.  Additionally, we have also  considered for this
reaction the rate given by  \citet[CF88]{cf:1988}. The comparison of all rates
is shown in Figure~\ref{f:o17rate}.

\begin{figure} 
\centerline{\includegraphics[scale=0.62]{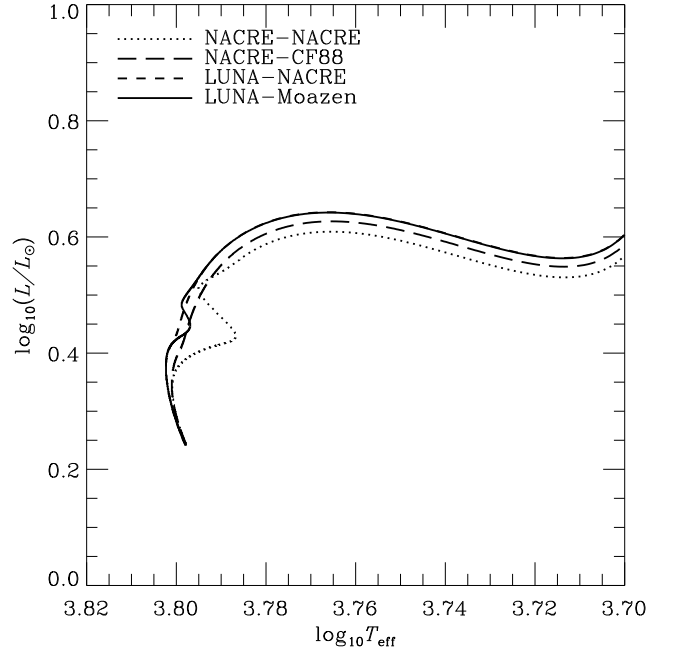}}
\caption{Evolution of a $1.2\,M_\odot$ star with GS98 composition, 
neglecting diffusion and overshooting, and using
  different rates for the two bottleneck reactions of the CNO-cycle (see text
for details). In the figure, the legends denote the source for the reaction
rates: left refers to the $\Iso{14}{N}(p,\gamma)\Iso{15}{O}$ reaction,
bottleneck of the whole CNO-cycle; right refers to the
$\Iso{17}{O}(p,\alpha)\Iso{14}{N}$ reaction, bottleneck of the CNO-II cycle.
\label{f:o17track}}
\end{figure}

In Figure~\ref{f:o17track}  we show the evolutionary track  of a 1.2~M$_\odot$
stellar model  with GS98 composition for different choices of  the two key
nuclear reactions mentioned above. The track corresponding to our standard
choice of nuclear rates, NACRE, is shown by the dotted line. Now, if the 
CF88 rate for the $\Iso{17}{O}(p,\alpha)\Iso{14}{N}$
reaction is used, which is less than 10\% of the NACRE rate 
throughout the relevant temperature regime, the track for this crucial mass is
strongly modified and no longer shows a sign of a convective core, as depicted
by the long-dashed line. If the 
\citet{moaz:07} rate is used, the  differences with respect to the track using
NACRE rates are very small; for this reason, we do not show this track. 
However, it is interesting to note that the effect that changing the 
$\Iso{17}{O}(p,\alpha)\Iso{14}{N}$ rate has on  the tracks depends on the rate
used for the 
$\Iso{14}{N}(p,\gamma) \Iso{15}{O}$ rate. Indeed, as stated before, using 
the LUNA rate for this reaction (and  NACRE rates for all others) leads to the
absence  of a convective  core in  the evolution  of a  $1.2\,M_\odot$ stellar
model.     This     is    shown     in     the     short-dashed    line     in
Figure~\ref{f:o17track}. However,
when in addition the \citet{moaz:07} rate  is used, a small convective core is
apparent in the track, as shown by the solid line in the same figure. 

We have  identified the reason  for this to  be the following: the  low
$\Iso{17}{O}(p,\alpha)\Iso{14}{N}$ rate  from CF88 slows down the CNO-II
subcycle, reducing thereby the flux of $\Iso{14}{N}$ back into the 
CNO-I cycle, which produces the overwhelming majority of the
energy. While the loss of energy from CNO-II is rather unimportant,
the drainage of available $\Iso{14}{N}$ by up to 90\% from the CNO-I
cycle is significant for its efficiency. One may consider this as a
storage of nitrogen in the form of useless $\Iso{17}{O}$. This makes
the appearance of a convective core more difficult, and thus raises
$M_\mathrm{ccc}$.      Note    that     for    the higher, older, 
$\Iso{14}{N}(p,\gamma)\Iso{15}{O}$ rate from NACRE this effect is not 
important; it appears that in this case the branching is favoring the
CNO-I cycle in any case, and the CNO-II cycle is always negligible.

Since the importance of the $\Iso{17}{O}(p,\alpha)\Iso{14}{N}$ was
rather surprising, and to make sure that our H-burning network included
all necessary reactions, we did a test calculation with a variant of
our code (Alves-Cruz 2009, private communication) that includes an
extensive p-capture network up to silicon. In particular, the
$\Iso{17}{O}(p,\gamma)\Iso{18}{F}$ reaction, which is the branching
reaction to the next higher cycle, is included. 
We saw no difference in the
evolution of stars in this mass range up to post main-sequence
phase. Our treatment of the CNO-cycles thus appears to be complete and
correct.

\subsection{Comparisons with other codes}
\label{s:codecomp}

We have seen that the behavior of the convective core for masses around the TO
mass in  M67 is  very sensitive  to the constitutive  physics included  in the
stellar models. Given  this sensitivity, it is important  also to determine if
calculation  done with  different stellar  evolution  codes lead  to the  same
conclusions.  Here we  have used,  in addition  to GARSTEC,  results  from two
additional codes:  LPCODE and  the Dartmouth code.  We summarize  our findings
below.

\begin{figure} 
\centerline{\includegraphics[scale=0.6]{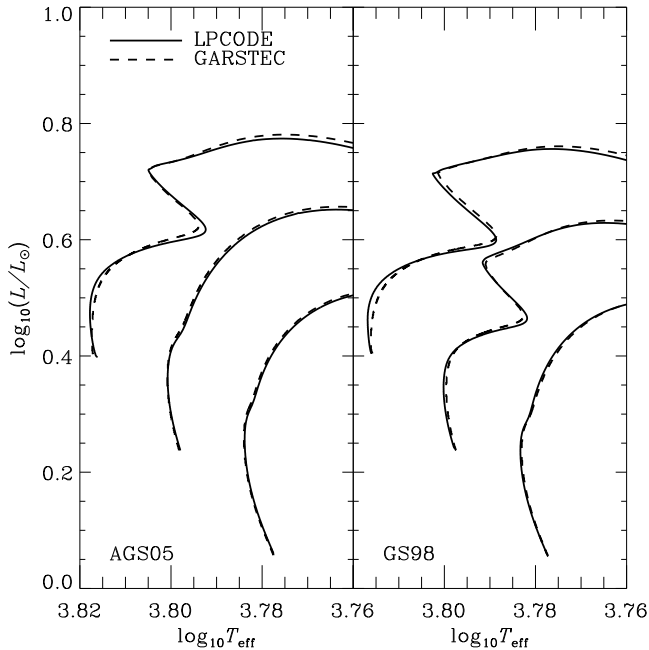}}
\caption{Comparison between LPCODE and GARSTEC.  Here the tracks  
were computed with NACRE and the updated $\Iso{14}{N}(p,\gamma)\Iso{15}{O}$ 
reaction  rate by \citet{mfn14:2008}, diffusion, and
overshooting, limited by the 
geometrical cutoff (Equation~\ref{eqn:ove2}) and a parameter of 
$f=0.018$.  Left  panel: models  with  AGS05  composition;  right panel:  GS98
composition. In  both cases tracks  shown are  1.1,  1.2, and
1.3~M$_\odot$, which cover the critical masses for M67.}
\label{f:laplata}
\end{figure}

LPCODE has originally been developed at La Plata Observatory. Extensive
descriptions of the code have been presented in \citet{alt:2002,alt:2003}.
Here we only describe updates and changes done in the code that were 
specifically implemented for this work.  The equation of state has been 
updated to the latest release by OPAL\footnote{http://adg.llnl.gov/Research/
OPAL/EOS\_2005/}. Radiative opacities from OPAL were calculated for both 
GS98  and AGS05 compositions and complemented at low temperatures with 
those from \citet{af:05}. Conductive opacities, although not relevant for this
work, 
are now those from Potekhin and collaborators, presented in \citet{condu:2007}. 
For the nuclear reaction rates, NACRE is the standard choice but the LUNA 
rate for the ${\rm^{14}N(p,\gamma)^{15}O}$ from \citet{mfn14:2008} has been 
adopted when  necessary. Overshooting is  treated in a diffusive  approach but
the geometric cutoff (Equation~\ref{eqn:ove2}) has been implemented to allow 
better comparison with GARSTEC results.

We  have  computed  a  large  set  of models  covering  the  ranges  of  mass,
composition and
input physics relevant to our work. In all cases, agreement between GARSTEC 
and   LPCODE   has   been   very   good.   Here,  we   choose   to   show   in
Figure~\ref{f:laplata} 
some evolutionary tracks that include  the most complex input physics, as used
for the CMD in in left panel of Figure~\ref{f:calibn14}: NACRE rates updated by
the 
$\Iso{14}{N}(p,\gamma)\Iso{15}{O}$  reaction rate from  \citet{mfn14:2008},
diffusion in  both the solar calibration  and in the  tracks, and overshooting
limited  by our  geometrical restriction  (Equation~\ref{eqn:ove2}).  Left and
right panels  show results for  AGS05 and GS98 compositions,  respectively. In
both cases, tracks for 1.1, 1.2, and 1.3~M$_\odot$ are shown, that bracket the
TO mass of  M67 and, consequently, determine the  morphology of the isochrones
around the TO.  
By inspection of Figure~\ref{f:laplata} it  can be readily seen that models with
both codes show a very similar evolution along the HRD; differences are hardly
noticeable. Results are of a similar quality for other choices of constitutive
physics and for both composition options. 
In relation to  systematic uncertainties affecting the conclusions
of  our work,  this result  is  particularly encouraging  because GARSTEC  and
LPCODE have been developed completely independently, and do not
share any numerical algorithm for solving the equations of stellar evolution.

\begin{figure}
\centerline{\includegraphics[scale=0.60]{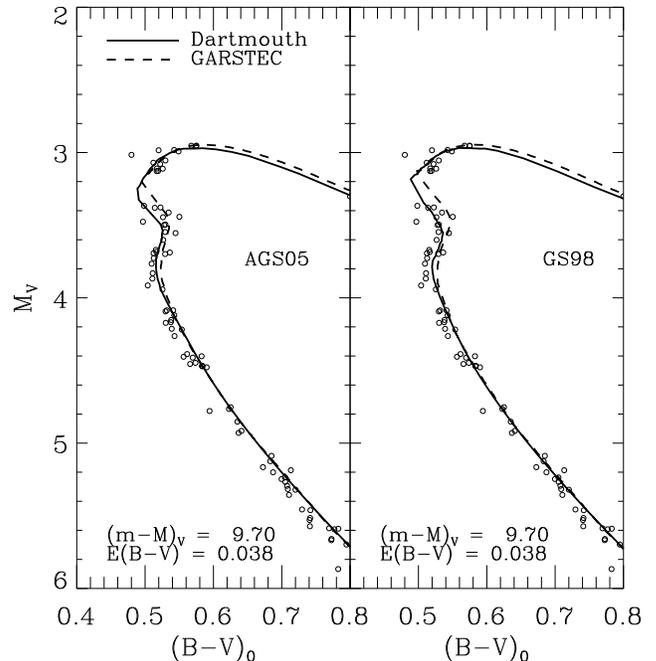}}
\caption{Comparison between the Dartmouth code and GARSTEC. CMDs of M67 showing,
for AGS05 and GS98 compositions, isochrones from both stellar evolution codes
including overshooting, diffusion and the LUNA rate for the
$\Iso{14}{N}(p,\gamma)\Iso{15}{O}$ reaction (see text for more details).
For each composition, morphology around the TO is similar for both codes. For
AGS05, the derived age is 4.5~Gyr, while for GS98 is 4.2~Gyr
(for both codes). \label{f:dartm}}
\end{figure}

The Dartmouth stellar evolution code is derived from the Yale code
\citep{gdpk:1992}, with modifications and updates described in
\citet{cfnps:2001}, \citet{bc:2006} and \citet{dcjkbf:2008}.
For this project, the NACRE nuclear reaction rates were implemented,
with the exception that the LUNA rate \citep{mfn14:2008} for the
$\Iso{14}{N}(p,\gamma)\Iso{15}{O}$ reaction was used. Convective core overshoot
in the  Dartmouth code is parametrized  as a multiple of  $H_P$. Normally, the
prescription of \citet{dwky:2004} 
is used, in which the amount of core overshoot is small
($0.05\, H_P$) for stars with small convective core masses, and
gradually increased to 0.20 $H_P$ for stars with masses
$0.2\,\mathrm{M}_\odot$ above the critical mass for the turn-on of
convection in the core. This prescription for convective core
overshoot was found to yield good agreement with open cluster
color-magnitude diagrams, when the \citet{gns:1996} solar mixture
was used. This mixture has $(Z/X)_\odot = 0.0244$. For this project,
a small constant amount of convective core overshooting regardless of 
the stellar mass was used. The amount of convective core overshoot used
in specific models is identified in the figure captions, and was
typically 0.07~$H_P$. 

In the case of the comparison between results of GARSTEC and the Dartmouth code
we choose to present CMD fits to M67 computed with both AGS05 and GS98
compositions. Models for GARSTEC use overshooting constrained by
Equation~\ref{eqn:ove3}. Results for the best fit isochrones for both codes are
shown in Figure~\ref{f:dartm}, where the left (right) panel shows results for
the AGS05 (GS98) composition. As throughout this paper, our goal is to
determine whether isochrones based on stellar models with one or the other
composition can reproduce the occurrence of the observed hook in M67.
As can be seen in Figure~\ref{f:dartm}, isochrones from both codes reproduce
well the TO morphology for both compositions. In the case of AGS05, the age of
for M67 obtained from both codes is 4.5~Gyr and for GS98 is 4.2~Gyr. Although
the TO morphologies are not identical for both codes, some differences are
likely to be present since we have not attempted, even within the uncertainties
given by the input physics and observational data, to obtain the best possible
agreement between the codes and with the data. 

As a last comparison, we state that the results determined
in VG07 (see \S~\ref{s:repro}) could also be reproduced by both codes. 

From comparing results from GARSTEC with those from LPCODE and Dartmouth code,
we conclude that, even if numerical schemes for solving stellar struture and
evolution equations and implementation of physics are different in the
different codes, our conclusions are robust; 
they do not depend on the stellar code used. It implies that systematics
between the codes are not an important source of uncertainty in our conclusions.

\section{Discussion and Conclusions}
\label{s:conc}

The open  cluster M67  has solar metallicity  and its  CMD shows a  clear hook
around  the  TO,  evidence  that  stars  populating  the  TO  have  convective
cores. The  mass of  these stars  is around 1.2~M$_\odot$,  very close  to 
the critical mass  value for which a convective core developes  as a result of
the efficient operation of  the CNO-cycle. Based on this,  VG07 have suggested
that M67 can  be used to test the new revisions  of solar abundances presented
by  Asplund and collaborators  \citep{ags:2005,ags:2009}, since  CNO elements,
catalysts in the CNO-cycle, 
have suffered the largest reductions in their solar abundance values. 
Here, we have reconsidered the viability of using M67 as a benchmark
for solar abundances, by studying  in detail under which conditions stars with
masses around 1.2~M$_\odot$ develop (or not) a convective core towards the end
of the  main sequence and how  this affects the morphology  of isochrones best
fitting M67 CMD.

We  have first  focused on  reproducing the  results obtained  by  VG07. Using
initial compositions for both AGS05  and GS98 obtained from calibrating solar
models without  element diffusion, we  have successfully recovered  their basic
results: if stellar models include a small amount of overshooting but no element
diffusion,  then the hook in M67 is  not reproduced by isochrones
computed with  the AGS05 solar  composition, while it  is present if  the GS98
composition is used  since, in this case, stars populating the  TO have a well
developed convective core. This has been  presented by VG07 as support for the
GS98 solar  composition; although  VG07 pointed out  to some caveats  in their
models that may influence this conclusion. 

We  then included  element diffusion  in our  models. On  the one  hand, solar
models   calibrated   with   diffusion   yield   a   higher   solar   initial
metallicity,  as summarized in  Table~\ref{t:ssm}. Additionally, because
gravitational settling  is the dominant  effect, metals tend to  accumulate in
the core. Inclusion of diffusion has then a double effect towards compensating
lower metallicities. Our  models with diffusion,  but still  using NACRE
nuclear rates, lead to the formation of  a convective core for TO stars in M67
even  with  the AGS05  composition  and,  consequently,  the distinctive  hook
observed  in  the  CMD  is  also  present  in  the  isochrones,  as  shown  in
Figure~\ref{f:ssmdiff}.  Under  these assumptions,  we find no  clear evidence
favoring the older, higher, GS98 solar metal abundances over AGS05.  

In our study, we have also identified two nuclear reactions, 
$\Iso{14}{N}(p,\gamma)\Iso{15}{O}$   and   $\Iso{17}{O}(p,\alpha)\Iso{14}{N}$,
that are
critical in determining the precise value of $M_\mathrm{ccc}$. The first
reaction is the bottleneck of the whole CNO-cycle, while the second one is the
bottleneck of the CNO-II part. When the temperature is not high enough,
the latter acts as a {\em  sink} for $\Iso{14}{N}$ nuclei, thereby slowing down
the whole 
CNO-cycle.  Let us first  discuss results  related to  the first  reaction. In
addition to the NACRE rate, we also used for 
$\Iso{14}{N}(p,\gamma)\Iso{15}{O}$  the latest measurement  by the  LUNA group
\citep{mfn14:2008}, which is about a factor of 2 lower than the NACRE rate.
As can be expected, the LUNA rate shifts
$M_\mathrm{ccc}$ to  larger values, by about 0.08~M$_\odot$.  With this rate,
even models with GS98 composition (not including diffusion) lack the
characteristic hook observed in  M67 (Figure~\ref{f:vg07n14}). This is not 
surprising in the
light of results with AGS05  composition, since the effective operation of the
CNO-cycle depends on the product of the abundance of catalysts and the 
$\Iso{14}{N}(p,\gamma)\Iso{15}{O}$ rate. The effect on the CNO-cycle
of reducing the rate for
this reaction by about a factor of 2, for stars with masses close to
$M_\mathrm{ccc}$ is very similar as to reducing the total number of catalysts
by a similar amount. A somewhat more unexpected result
relates  to  the   second  reaction,  $\Iso{17}{O}(p,\alpha)\Iso{14}{N}$.  The
branching  between  the CNO-I  and  CNO-II  cycles  is almost  independent  of
temperature for the temperature range that interests us in this
work.  However, if  the temperature  is not high enough for  the CNO-II cycle 
to be fully active,  $\Iso{17}{O}$ is  created at  the  expense of 
$\Iso{14}{N}$ but  the
feedback  to the  CNO-I cycle (the  only  one relevant  in terms  of
energetics) is
inefficient. As a consequence, the  total number of catalysts is reduced, with
$\Iso{14}{N}$ being stored in the form of $\Iso{17}{O}$.  
For this rate we  have used, in addition to the NACRE  rate, the older rate by
\citet{cf:1988} and the new measurement by \citet{moaz:07}. 
When we use the \citet{cf:1988} rate, that is about 10\% of the NACRE rate
in the relevant temperature range (see Figure~\ref{f:o17rate} for a comparison
of the rates), we find that, even for GS98 and the NACRE
rate for  $\Iso{14}{N}(p,\gamma)\Iso{15}{O}$ there is no convective  core in a
1.2~M$_\odot$ models  towards the  end of the  main sequence.  This translates
into the absence of a hook in  the isochrones that would best fit M67 CMD. One
can regard the last exercise as merely of academic interest since the low rate
from  \citet{cf:1988}  seems  now to be ruled  out  by  experiments.  However, 
the importance  of this  rate is  still worth  being taken  into account  with
the newest results by \citet{moaz:07}, even  if this rate agrees with NACRE
within 40\%.   By  using  the \citet{moaz:07}  rate  in   the  models,  the
1.2~M$_\odot$ model  with the  LUNA rate for the
$\Iso{14}{N}(p,\gamma)\Iso{15}{O}$ reaction recovers a small convective core. 
These results are summarized in Figure~\ref{f:o17track}. 

The development of convective cores in stars close to $M_\mathrm{ccc}$ is
sensitive to the detailed physical input in the models, as seen from the above
results. Moreover, an appreciable amount  of freedom in the modeling is still
present because  of the  lack of  a proper convection  theory to  account for,
particularly,  the amount  of overshooting  occurring in  stellar cores.  It is
widely accepted, however, that an overshooting region of about 0.2~$H_P$
is  needed in stars  with masses  above $\sim  2$~M$_\odot$ to  explain, among
others, 
the width of main sequences in  stellar clusters. On the other hand, for stars
below that mass, and particularly close to $M_\mathrm{ccc}$, it is also known
that overshooting has  to be limited to a smaller  region. However, a detailed
understanding of  how core overshooting depends of  stellar parameters, namely
the stellar mass, is not known. We have used here two different approaches to
limit  overshooting,   a  geometric  cutoff   and  an efficient factor that
increases linearly with stellar
mass, Equations~\ref{eqn:ove2}~and~\ref{eqn:ove3} respectively, with  the
geometric
cutoff  being the  standard  choice.  In relation  to  core overshooting,  its
importance is that 
it helps to reproduce the hook  in M67 because it facilitates stars to sustain
a convective core during the main sequence and the growth of the core
towards  the end  of the  main sequence  for stars with  masses around
$M_\mathrm{ccc}$, critical for explaining the TO morphology in M67. 
On the other  hand, overshooting cannot, by itself,  {\em create} a convective
core when there  is none.  This is  the case, for example, for our models with
AGS05 composition and the LUNA  rate for the $\Iso{14}{N}+p$ reaction when the
geometric cutoff is applied, as shown in the left panel of
Figure~\ref{f:calibn14}.  In this case,  the convective core developed towards
the end of the 
pre-main sequence evolution cannot be  sustained by stars with masses close to
the TO mass in M67, even with the help of overshooting. This may be the result
of our geometric cutoff being {\em too restrictive}. When, instead, the linear
ramp function  of Equation~\ref{eqn:ove3} is  used, isochrones based  on AGS05
show   again    the   characteristic   hook    of   M67   (right    panel   in
Figure~\ref{f:calibn14}).   With  this   prescription  for  overshooting,  for
stellar masses around  M67 TO masses, the convective  core survives during all
main sequence 
evolution and  naturally grows  towards the  end of it,  due to  increased CNO
burning, leading to the occurrence of a hook in M67 CMD.
Overshooting provides a certain degree  of freedom in stellar models, and here
we show that this freedom renders isochrone fitting to 
M67 uncertain,  in that depending on  the way overshooting is  modeled, we are
able to find, or not, a hook around  the TO. It is important to note that both
prescriptions for the limitation of overshooting for stellar masses around 
$M_\mathrm{ccc}$ can be made consistent with current observational evidence. 

We have  also tried to assess the possible  dependence of our results
on the numerics, that is the  stellar evolution code. Therefore, we have
used,  in addition  to  GARSTEC, two  other  evolution codes:  LPCODE and  the
Dartmouth  code.  None of  these  codes share  a  common  background in  their
development. For  this work, when possible,  we have tried to  match the input
physics, but some  differences still remain. The most  important, probably, is
that in the Dartmouth code overshooting  is accounted for as a fraction of the
pressure scale height. For models  appropriate for M67, this fraction has been
taken as 0.07~$H_P$. Comparison of results between the three codes
is   highly   satisfactory.   Solar   model   callibrations,   summarized   in
Table~\ref{t:ssm},  yield   very  small  differences.   A  comparison  between
evolutionary   tracks  computed   with  LPCODE   and  GARSTEC   is   shown  in
Figure~\ref{f:laplata}. The  agreement between results  from the two  codes is
excellent. A comparison of isochrone fitting to the CMD of M67 with GARSTEC and
the Dartmouth code is shown in Figure~\ref{f:dartm} for both compositions used
in this work. Again, results are similar with both codes, which are able to
reproduce the observed morphology around the TO. For each composition, both
codes also yield the same age for the cluster, 4.5 and 4.2~Gyr for AGS05 and
GS98 compositions respectively. We conclude from these tests that systematics
originating in the use of different numerical schemes, and even some differences
in the implementation of the input physics, do not affect our conclusions.

Finally, we comment on two additional points that reinforce our main
conclusions given below. As stated in \S~\ref{s:data}, the uncertainty in the
metallicity of M67, according to \citet{gm67:2000} is about 0.06~dex, i.e.
15\%. We have not played with the metallicity of our models and have always
assumed a solar value (where solar means GS98 or AGS05). However, in view of the
dependence of the precise value of $M_\mathrm{ccc}$ to the details
of the constitutive physics, variations in the metallicity of our
models can in principle affect the occurrence or not of a convective core for
stars with M67 TO-mass. 
The last point is that in the new determination of
solar abundances by \citet{ags:2009} CNO elements have been revised
upwards. The total number of CNO catalysts in the new \citet{ags:2009} is about
8\% higher than in the previous AGS05 solar abundance compilation. This increase
would also contribute to facilitate the occurrence of a convective core at the
TO-mass in M67. 

The main conclusions we draw from our study are as follow. The different solar
compositions, namely AGS05 and GS98  certainly have an impact on the predicted
morphology of  the CMD in M67. The occurrence of the TO-hook is more difficult
to achieve for the low CNO abundances of AGS05. This  is a confirmation of 
previous results by VG07. However, we  have also  found
that  other constitutive physics  in the
models,  e.g.  element  diffusion,  nuclear reactions,  prescription  of  core
overshooting, also influence the stellar  mass at which convective cores start
to develop. This translates, in the case of M67, into isochrones that may have
a hook around the TO even with the AGS05 composition or, on the contrary, that
show  no  hook  even  for  the  GS98 composition.  There  is  a  certain  {\em
  degeneracy} in the constitutive physics  underlying the presence of the hook
in   M67  that  cannot   be,  at   the  present   status  of   our  knowledge,
disentangled. The morphology  of M67 could in principle be  used to test solar
abundances,  but only  under  the  strong assumption  that  all other  factors
affecting the TO morphology, more precisely the occurrence of a convective
core in stars with masses around the TO-mass, are completely under control.
Since we understand this is not the  case, we conclude that M67 CMD morphology 
does not present a strong argument against low CNO abundances in the Sun.

\acknowledgments

We wish to thank Don A. Vandenberg for his very helpful input
to this work and his openess to discuss our results and compare
them with his own. M. Alves-Cruz did additional calculations,
which we gratefully acknowledge. We also thank
M. Asplund, S. Cassisi and L. Pasquini for helpful discussions
and advice.

\end{document}